\documentclass{aastex63}
\usepackage{amsmath}
\usepackage{units}
\usepackage{gensymb}

\begin{document}
\title{Exploring the stellar rotation of early-type stars in the LAMOST
Medium-Resolution Survey. I. Catalog}	

\author[0000-0002-3279-0233]{Weijia Sun} 

\affiliation{Kavli Institute for Astronomy \& Astrophysics and
  Department of Astronomy, Peking University, Yi He Yuan Lu 5, Hai
  Dian District, Beijing 100871, China}
\affiliation{Key Laboratory for Optical Astronomy, National
  Astronomical Observatories, Chinese Academy of Sciences, 20A Datun
  Road, Chaoyang District, Beijing 100012, China}
  
\author[0000-0002-6573-6719]{Xiao-Wei Duan}
\affiliation{Kavli Institute for Astronomy \& Astrophysics and
  Department of Astronomy, Peking University, Yi He Yuan Lu 5, Hai
  Dian District, Beijing 100871, China}
\affiliation{Key Laboratory for Optical Astronomy, National
  Astronomical Observatories, Chinese Academy of Sciences, 20A Datun
  Road, Chaoyang District, Beijing 100012, China}

\author[0000-0001-9073-9914]{Licai Deng}
\affiliation{Key Laboratory for Optical Astronomy, National
  Astronomical Observatories, Chinese Academy of Sciences, 20A Datun
  Road, Chaoyang District, Beijing 100012, China}
\affiliation{School of Astronomy and Space Science, University of the
  Chinese Academy of Sciences, Huairou 101408, China}
\affiliation{Department of Astronomy, China West Normal University,
  Nanchong 637002, China}
  
\author[0000-0002-7203-5996]{Richard de Grijs}
\affiliation{Department of Physics and Astronomy, Macquarie
  University, Balaclava Road, Sydney, NSW 2109, Australia}
\affiliation{Research Centre for Astronomy, Astrophysics and
  Astrophotonics, Macquarie University, Balaclava Road, Sydney, NSW
  2109, Australia}
  
\author[0000-0002-6434-7201]{Bo Zhang}
\affiliation{Department of Astronomy, Beijing Normal University,
  Beijing 100875, China}

\author[0000-0002-1802-6917]{Chao Liu}
\affiliation{Key Laboratory for Optical Astronomy, National
  Astronomical Observatories, Chinese Academy of Sciences, 20A Datun
  Road, Chaoyang District, Beijing 100012, China}

\begin{abstract}
We derive stellar parameters and abundances (`stellar labels') of
40,034 late-B and A-type main-sequence stars extracted from the Large
Sky Area Multi-Object Fiber Spectroscopic Telescope Medium Resolution
Survey (LAMOST--MRS). The primary selection of our early-type sample
was obtained from LAMOST Data Release 7 based on spectral line
indices. We employed the Stellar LAbel Machine (SLAM) to derive their
spectroscopic stellar parameters, drawing on Kurucz spectral synthesis
models with $\unit[6000]{K} < T_\mathrm{eff} < \unit[15,000]{K}$ and
$\unit[-1]{dex} < \mathrm{[M/H]} <\unit[1]{dex}$. For a
signal-to-noise ratio of $\sim 60$, the cross-validated scatter is
$\sim\unit[75]{K}$, \unit[0.06]{dex}, \unit[0.05]{dex}, and
$\sim\unit[3.5]{km\,s^{-1}}$ for $T_\mathrm{eff}$, $\log g$, 
[M/H], and $v\sin i$, respectively. A comparison with objects with
prior, known stellar labels shows great consistency for all stellar
parameters, except for $\log g$. Although this is an intrinsic caveat
that comes from the MRS's narrow wavelength coverage, it only has a
minor effect on estimates of the stellar rotation rates because of the
decent spectral resolution and the profile-fitting method
employed. The masses and ages of our early-type sample stars were
inferred from non-rotating stellar evolution models. This paves the
way for reviewing the properties of stellar rotation distributions as
a function of stellar mass and age.
\end{abstract}

\keywords{Stellar rotation (1629), Astronomy data analysis (1858),
  Catalogs (205), Early-type stars (430), Stellar properties (1624)}
  
\section{Introduction}
\label{sec:intro}

Early-type stars comprise hot, massive, and luminous stars of spectral
types O, B, A, and early-F. They are short-lived and rare compared
with late-type dwarfs, which are less massive. Most early-type stars
are found in dense interstellar clouds in spiral arms or other dusty
regions. As massive stars, they facilitate the chemical enrichment and
reionization of the Universe, which makes them important contributors
to the evolution of their host galaxies.

Early-type stars are very important for studies of Galactic and
stellar structure. They are the progenitors of various types of
supernovae \citep{2008ApJ...675..614P} and can be sources of gamma-ray
bursts \citep{2012ARA&A..50..107L}. During their lifetimes, they
release much ultraviolet radiation, which rapidly ionizes the
surrounding interstellar medium in giant molecular clouds, forming
H{\sc ii} regions or Str{\"o}mgren spheres. Massive early-type stars
are useful probes of elemental abundances and tracers for mapping the
spiral structure and disk of the Milky Way \citep{1952AJ.....57....3M,
  1995ApJS...99..659V, 2010ApJ...718..683C, 2017AJ....153...99C,
  2018A&A...616L..15X}.

Most early-type stars have significantly higher rotation rates than
solar-type and low-mass stars \citep{2000AcA....50..509G}, due to
their higher initial angular momenta, shorter contraction timescales
to the zero-age main sequence (ZAMS), lack of deep convective
envelopes, and strong magnetic fields (except for chemically peculiar
stars). For instance, Be stars are B-type main-sequence stars with
Balmer-line emission \citep{2013A&ARv..21...69R}. The latter is
believed to originate from their decretion disks, fed by mass ejected
from the central stars due to rapid rotation (up to
$\unit[\sim350]{km\,s^{-1}}$). Binarity is another key parameter that
is different between early-type stars and their late-type
counterparts. The overall multiplicity frequency of main-sequence
stars is a steep, increasing, monotonic function of spectral type from
late to early types \citep{2013ARA&A..51..269D}.

Precise and accurate stellar parameters derived from stellar spectra
are necessary to understand early-type stars. Thanks to rapid
improvements in modern observing facilities, we can now access large
numbers of spectra from large spectroscopic surveys, e.g. the Large
Sky Area Multi-Object Fiber Spectroscopic Telescope survey
\citep[LAMOST;][]{2012RAA....12..735D}, the Sloan Digital Sky Survey
\citep[SDSS;][]{2011AJ....142...72E}, the Apache Point Observatory
Galactic Evolution Experiment \citep[APOGEE;][]{2017AJ....154...94M},
or the \textit{Gaia} Radial Velocity Spectrometer
\citep{2018A&A...616A...5C}, which provide unprecedented opportunities
to unravel the stellar population properties of large stellar
samples. \cite{2019ApJS..241...32L} presented 22,901 stellar spectra
of OB stars identified in LAMOST Data Release (DR) 5 based on their
distribution in the space defined by their spectral line indices. A
similar approach was applied by \citet{2021arXiv210909775G}. These authors
reported 10,608 early-type stars from LAMOST DR7 and estimated the
binary fractions of different spectral types.

To derive stellar `labels' (stellar parameters and abundances), common
practice is to match stellar spectra and model templates using Balmer
and metal lines. In recent years, machine-learning algorithms have
been introduced, which have proved efficient in processing large
databases
\citep{2019ApJ...879...69T,2019ApJS..245...34X}. \cite{2020ApJS..246....9Z}
reported the implementation of a data-driven method based on support
vector regression (SVR), known as the Stellar LAbel Machine (SLAM).
It can produce stellar labels for a wide range of spectral types in
LAMOST DR5 with high efficiency.

Here, we undertake a survey of early-type stars in LAMOST DR7, based
on a sample size that is more than ten times larger than the
previously largest database \citep{2012A&A...537A.120Z}. We construct
a data-driven model for medium-resolution spectra based on the SLAM
model trained by ATLAS12 atmospheric models
\citep{2005MSAIS...8..189K}. We compile a catalog composed of more
than 40,000 stars with effective temperatures of $\unit[7000]{K} \leq
T_\mathrm{eff} \leq \unit[14,500]{K}$ and derive their stellar labels
($T_\mathrm{eff}$, $\log g$, [M/H], and $v\sin i$) and
fundamental parameters (masses and ages).

This article is organized as follows. In Section~\ref{sec:data}, we
introduce the primary selection of our early-type candidates from the
LAMOST data. Our application of the SLAM method to derive stellar
labels, as well as the results of our validation, are presented in
Section~\ref{sec:method}. The final catalog, including estimates of
the fundamental stellar parameters and possible contamination, is
described in Section~\ref{sec:results}. Finally, conclusions are drawn
in Section~\ref{sec:conclusions}.

\section{Data}
\label{sec:data}
\subsection{Primary selection}
\label{sec:cat}

LAMOST is a 4-meter quasi-meridian reflective Schmidt telescope
equipped with 4000 fibers across its $5^\circ$ field-of-view focal
plane \citep{2012RAA....12.1197C, 2012RAA....12..723Z}. Following the
initial low-spectral-resolution ($R \sim 1800$) survey (LRS),
conducted from 2011 to 2018 \citep[LAMOST-I;][]{2015RAA....15.1095L},
LAMOST has been conducting a medium-resolution survey (MRS), $R \sim
7500$, with its upgraded spectrographs
\citep[LAMOST-II;][]{2020arXiv200507210L}. The blue cameras of each
spectrograph cover the wavelength range from \unit[4950]{\AA} to
\unit[5350]{\AA}, while the red cameras cover \unit[6300]{\AA} to
\unit[6800]{\AA}. With a single exposure time of \unit[1200]{s}, the
median signal-to-noise ratio (SNR) can reach values in excess of 5 for
stars with \textit{Gaia} $G$-band magnitudes $G <
\unit[14.5]{mag}$. The SNR of co-added spectra composed of three
\unit[1200]{s} spectra can reach 10 at $G=\unit[14.5]{mag}$.
 
MRS DR7 contains 11,422,346 single-exposure and 2,968,667 co-added
spectra (encompassing both the blue and red bands), which are
tabulated in the LAMOST MRS General Catalog. The LAMOST data
reduction pipeline subtracts the sky background and removes the
Earth's atmospheric telluric absorption in four bands, including the
$B$ band \citep{2013IAUS..295..189Z}. The LAMOST team also released
the LAMOST MRS Parameter Catalog, a subsample of 807,319 spectra whose
stellar parameters and chemical abundances were estimated using
various pipelines. The effective temperatures, surface gravities, and
metallicities were determined by the LAMOST Stellar Parameter pipeline
(LASP), whereas the rotational velocity was measured by matching with
ELODIE templates. The elemental abundances were estimated using a
deep-learning method \citep{2020ApJ...891...23W}.

To compile our primary catalog of early-type ($T_\mathrm{eff} >
\unit[7000]{K}$) candidates, we selected the co-added spectra from the
LAMOST MRS General Catalog and cross-matched them with \textit{Gaia}
early DR3 \citep[EDR3;][]{2021A&A...649A...1G}. Compared with
\textit{Gaia} DR2 \citep[][]{2018A&A...616A...1G}, EDR3 represents a
significant improvement as regards estimates of the parallax
systematics, both globally \citep{2021A&A...649A...4L} and locally
\citep{2021ApJ...911L..20R}. \citet{2021A&A...649A...4L} published the
parallax zero-point based on their analysis of quasars, binary stars,
and the Large Magellanic Cloud. We adopted the photo-geometric
distance estimates of \citet{2021AJ....161..147B}---who inferred
distances from EDR3 parallaxes (corrected for the parallax zero-point)
based on a three-dimensional Galactic model---as the distances of our
candidates. We also cross-matched our candidates with the Two Micron
All Sky Survey \citep[2MASS;][]{2006AJ....131.1163S} for extinction
correction (see Section~\ref{sec:bol}). As our preliminary selection
criteria, we rejected those objects with effective temperatures below
\unit[6500]{K} (in practice below \unit[7000]{K} to avoid removal of
objects with incorrect temperature estimates) or SNRs less than 15.

Next, we selected those stars with accurate \textit{Gaia} astrometric
and photometric parameters following \citet{2021A&A...649A...3R}. The
renormalized unit weight error (RUWE), introduced after \textit{Gaia}
DR2 \citep{2018A&A...616A..17A}, was adopted as a goodness-of-fit
indicator ($\texttt{ruwe} < 1.4$). Objects with a close separation on
the sky, less than 0.7\arcsec \citep{2021A&A...649A...1G}, cannot
yield reliable measurements. Therefore, they were rejected by imposing
the criterion \texttt{duplicated\_source}$==$\texttt{True}. We further
used the corrected BP and RP flux excess factor $C^*$, introduced by
\citet{2021A&A...649A...3R}, to quantify the impact of variations in
the local background level. The latter may affect the $G_\mathrm{BP}$
and $G_\mathrm{RP}$ integrated photometry. Equation 18 of
\citet{2021A&A...649A...3R} was used to select a sample of
well-behaved, isolated stellar sources with good-quality
photometry. No saturation correction was applied, because the bright
stars in our sample do not reach the brightness levels where
saturation becomes an issue. As the final step, we rejected candidates
with relative parallax errors larger than 10\%.

\subsection{Line indices}

We further used the objects' spectral line indices to select
early-type candidates. Line indices provide a mapping of the effective
temperature without the caveats associated with extinction correction
and flux calibration \citep{2015RAA....15.1137L}. A line index, in
terms of its equivalent width (EW), is defined as
\citep{1994ApJS...94..687W}
\begin{equation}
	\mathrm{EW} =
\int\left[1-\frac{f_\mathrm{line}(\lambda)}{f_\mathrm{cont}(\lambda)}\right
]\mathrm{d} \lambda,
\end{equation}
where $f_\mathrm{line}(\lambda)$ and $f_\mathrm{cont}(\lambda)$ are
the fluxes of the spectral line and the continuum, respectively, and
$\lambda$ is the rest-frame wavelength.

Normalization of the spectra and transformation to the rest
frame was done using the
\texttt{laspec}\footnote{\url{https://github.com/hypergravity/laspec}}
package \citep{2021arXiv210511624Z}. In brief, we normalized our
spectra by iteratively fitting a spectrum three times with a spline
function and rejecting pixels deviating by more than 3$\sigma$ from
the median values in \unit[20]{\AA} windows. Following normalization,
the radial velocities, $v_{R}$, were estimated using the
cross-correlation function method with respect to a collection of
synthetic spectra based on the ATLAS9 model atmospheres
\citep{2018A&A...618A..25A}. Because of the narrow wavelength coverage
of both the blue and red spectrographs ($\sim\unit[400]{\AA}$),
continuum fluxes may be affected by the non-uniform spectral response
function, especially toward the edges. Therefore, we calculated the
relative flux ratios by normalization of the entire spectrum, instead
of using the pseudo-continuum \citep[e.g.][]{2015RAA....15.1137L,
  2021arXiv210909775G}. Note that there might be some offset between the blue and
red arms. Therefore, we applied the cross-correlation technique to
both arms and transformed the red and blue spectra to the rest
frame separately. We discarded those sources with $|v_{R}| >
\unit[250]{km\,s^{-1}}$ since they are probably affected by bad pixels
suffering from strong cosmic rays or contaminated by emission lines
arising either from their surrounding disks, or nearby H{\sc ii}
regions.

\begin{figure}[ht!]
\plotone{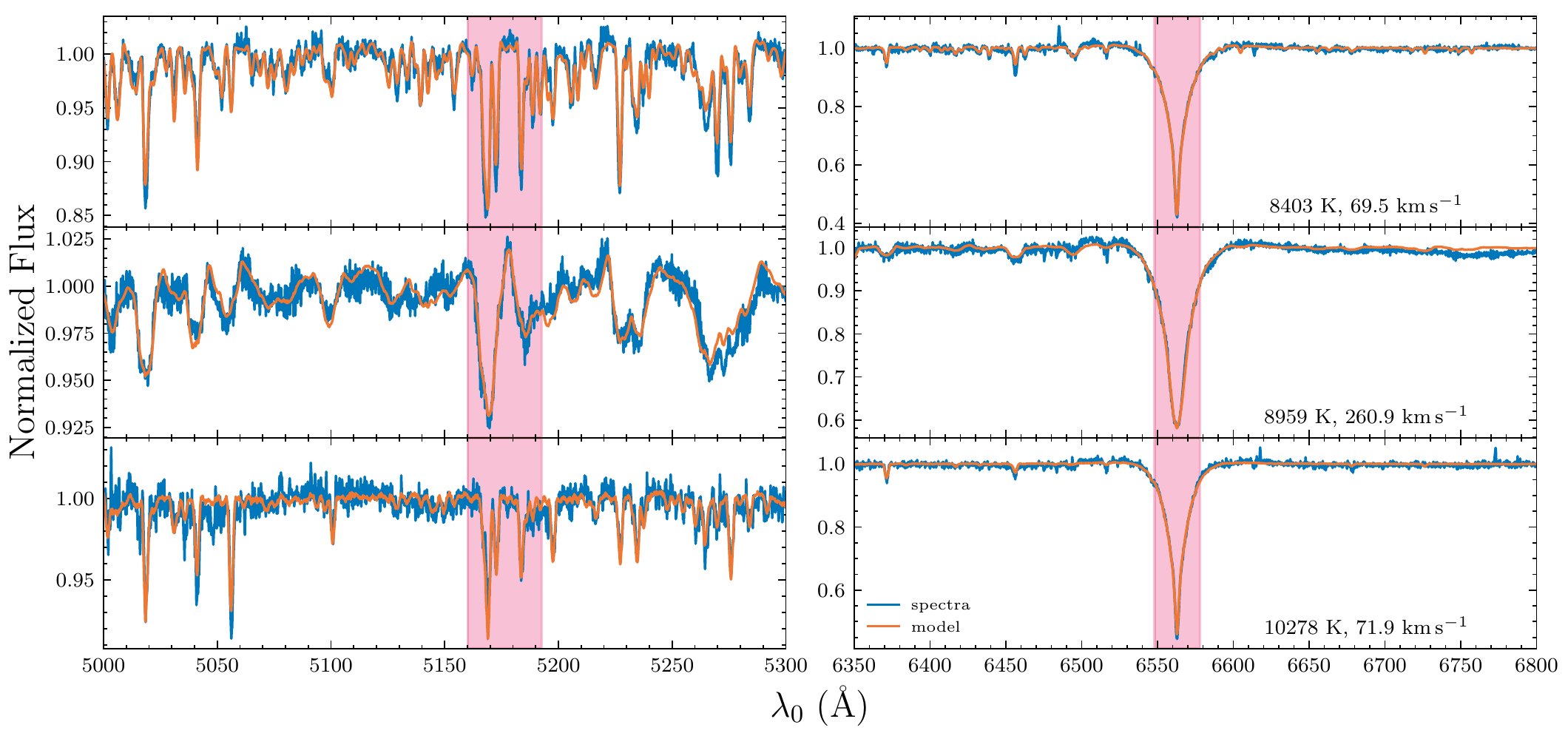}
\caption{Normalized sample spectra, in the rest frame, from
LAMOST MRS DR7 (blue) and the corresponding best-fitting models
(orange) determined by the SLAM algorithm (see
Section~\ref{sec:method}). Blue and red segments of the same
object are shown in the left and right panels, respectively. The
best-fitting effective temperatures, $T_\mathrm{eff}$, and the
projected rotational velocities, $v\sin i$, are listed in the
bottom right corners. The red bands correspond to the spectral
regions covering Mg\,{\sc i} \textit{b}
(\unit[5160.12--5192.62]{\AA}) and H$\alpha$
(\unit[6548.0--6578.0]{\AA}).
\label{fig:sample}}
\end{figure}

In Figure~\ref{fig:sample}, we present three sample spectra
characterized by different effective temperatures from LAMOST MRS
DR7. Observations of the same object obtained with the blue and red
arms are shown in the left and right panels, respectively. The
wavelength scale was transformed to the rest frame and fluxes were
normalized to the pseudo-continuum, as described above. The
best-fitting models determined by application of the SLAM method are
shown as orange curves. The vertical red bands represent the
wavelength ranges covering Mg\,{\sc i} \textit{b} and H$\alpha$,
which were used in our selection of the line indices. The top and
bottom panels represent spectra of two stars with small projected
rotational velocities ($v\sin i < \unit[100]{km\,s^{-1}}$), while
the middle panel shows a spectrum that is affected by strong
rotational broadening.

\begin{figure}[ht!]
\plotone{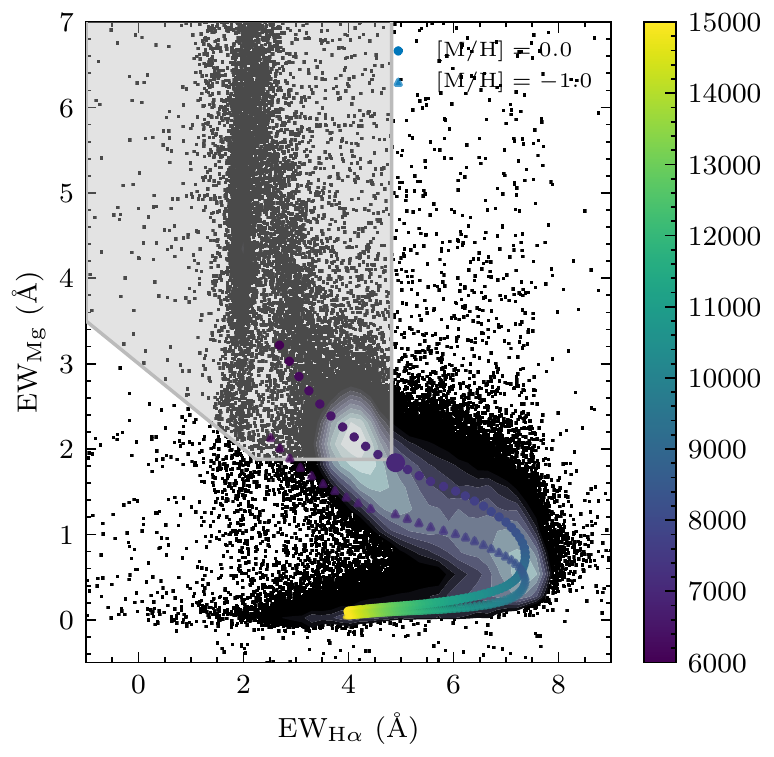}
\caption{Distribution of the H$\alpha$ and Mg\,{\sc i} \textit{b} line
  indices of the MRS sample. Regions with high number densities are
  shown as grayscale contours. Reference line indices from the ATLAS
  atmospheric models in the range from \unit[6000]{K} to
  \unit[15,000]{K} are overplotted as colored circles ($\log g =
    \unit[4.0]{dex}$, [M/H]$=\unit[0.0]{dex}$) and triangles
    ($\log g = \unit[4.0]{dex}$, [M/H]$=\unit[-1.0]{dex}$).
  The large circle at $\mathrm{EW}_\mathrm{Mg}\sim\unit[2]{\AA}$
  represents the model for $T_\mathrm{eff} = \unit[7000]{K}$. The
  shadowed area represents the selection boundary for stars cooler
  than $\unit[7000]{K}$. \label{fig:ew}}
\end{figure}

Figure~\ref{fig:ew} shows the stellar loci in the parameter space
defined by the line indices of H$\alpha$ (\unit[6548.0--6578.0]{\AA})
and Mg\,{\sc i} \textit{b} (\unit[5160.12--5192.62]{\AA}, reflecting integration over the three
  main Mg\,{\sc i} features at 5167, 5172, and 5183 {\AA}) for the
selected MRS sample objects. The contours show the number densities in
dense regions. Line indices from the ATLAS atmospheric models are
overplotted for reference, color-coded by their effective
temperature. The dense cluster at
$\mathrm{EW}_\mathrm{H\alpha}\sim\unit[4]{\AA}$ represents the
low-temperature sample objects ($T_\mathrm{eff} < \unit[6500]{K}$)
remaining following the initial temperature selection. There is an
elongated aggregation at
$\mathrm{EW}_\mathrm{H\alpha}\sim\unit[2]{\AA}$, which is mostly
composed of late-type stars with $T_\mathrm{eff} < \unit[3100]{K}$
that were missed in the LASP, while another aggregation is located at
bottom right around \unit[9400]{K}. The latter is an artifact caused
by a concentration in the line indices parameter space (see
Section~\ref{sec:bol}).

To further avoid contamination by late-type stars, we rejected
candidates with $T_\mathrm{eff} < \unit[7000]{K}$ (the shadowed area
in Figure~\ref{fig:ew}). Finally, we were left with 84,382 early-type
candidates with decent SNRs (with median SNR values in excess of
40). As the next step in our data processing, stellar labels
($T_\mathrm{eff}$, $\log g$, [M/H], and $v\sin i$) of each star
in this reduced sample were extracted. We will outline the adopted
method in the next section.

\section{Method}
\label{sec:method}

SLAM is a data-driven method to estimate precise stellar labels from
spectra based on SVR. Compared with other attempts, also based on SVR
\citep[e.g.][]{2015ApJS..218....3L}, SLAM applies a generative model
that automatically adjusts the model complexity and extracts
information from the stellar spectra. SLAM's performance on
application to the LAMOST LRS \citep{2020ApJS..246....9Z} and MRS
\citep{2020RAA....20...51Z} spectra has been inspected, and the
approach has been applied to various stellar populations
\citep[e.g.][]{2021arXiv211006246G, 2021ApJS..253...45L}.

\subsection{Training}
\label{sec:train}

Data-driven methods \citep[e.g.][]{2015ApJ...808...16N,
  2017ApJ...849L...9T} presume that the stellar labels of the training
set are known accurately and precisely. Ideally, the training set is
composed of observed spectra with high SNRs that have high-fidelity
labels, either from a reference set built from well-studied objects
that have been observed in the context of the survey
\citep{2015MNRAS.449.1401H} or from a common subset taken from a
higher-resolution survey \citep{2017ApJ...849L...9T}. Unfortunately,
such a training set containing a large collection of early-type
stellar labels is not available. Therefore, we used a set of mock
stellar spectra from ATLAS12 atmospheric models to build the training
set.

However, our stellar labels are referenced to the set of model
  spectra adopted. For instance, the effective temperatures derived
  for chemically peculiar stars---which are prevalent especially among
  A-type stars \citep{1974ARA&A..12..257P}---depend on which lines are
  used \citep[Ca\,{\sc ii} K, Balmer H, metallic
    lines;][]{2009ssc..book.....G}. Our sample probably also includes
  shell stars \citep[e.g., Pleione;][]{1977ApJS...35..441G}, which
  have narrow absorption lines associated with cooler plasma that
  originates in a gas disk projected against the stellar
  photospheres. In this paper, the results for such objects depend on
  which features are covered by the blue and red segments of the
  observed MRS spectra. Also note that the adopted ATLAS12 model is a
  local thermodynamic equilibrium (LTE) spectral model that does not
  take into account the presence of non-LTE (NLTE) effects. The
  atmospheres of cool stars are well described by LTE model
  atmospheres, but stars of spectral types earlier than A require a
  detailed treatment of NLTE effects because of the high energy
  densities of their radiation fields \citep{2011JPhCS.328a2015P}.
  \citet{2018ApJ...866..153A} presented NLTE models for the Mg\,{\sc
    i} \textit{b} lines, which is of importance in our sample's
  context since neutral magnesium is a minority species. They found
  that departures from LTE are essential for stars hotter than
  $\unit[9000]{K}$, potentially leading to differences in Mg abundance
  of several tenths of a dex. Although our method is inevitably
limited by the systematic shortcomings embedded in the theoretical
model spectra, this approach can perform well enough in practice
\citep{2017ApJ...843...32T}.

The training process involves two different SLAM models. The first
model is used to convert high-resolution spectra from the model grid
($T_\mathrm{eff}$, $\log g$, $\lbrack\mathrm{Fe}/\mathrm{H}\rbrack$)
to yield random values for these parameters within predefined ranges. The
second one is used to construct mock spectra resembling the MRS spectra using
synthetic spectra convolved with the rotational profile ($v\sin i$) and
instrumental broadening.

Note that the specific intensity profiles pertaining to the
  stellar atmosphere vary from center to limb. The convolution
  approach is an approximate representation of the rotation profile
  under the assumption that center-to-limb variations are negligible
  compared to rotation effects \citep{2005oasp.book.....G}. This
  approximation will cease to hold for rapid rotation, which induces
  temperature and gravity inhomogeneities, leading to differences that
  depend on the spin axis orientation with respect to the line of
  sight. Such gravitational darkening effects, most notable in
  Be---B-type stars exhibiting Balmer-line emission
  \citep{2004MNRAS.350..189T}---will result in an underestimation of
  the rotation velocities. Nevertheless, for the majority of our
  sample objects this effect is minor and only a small fraction of
  rapid rotators suffer from this systematic error. As for the limb
  darkening law, we adopted a fixed value for the linear
  limb-darkening coefficient, $\epsilon=0.6$, a value typical for the
  range of temperatures found among our sample
  \citep{2002A&A...381..105R}. \citet{1985A&AS...60..471W} pointed out
  that the coefficient depends on the effective temperature and the
  wavelength, and \citet{2011A&A...531A.143D} estimated that
  neglecting these variations may produce an error of $\sim2$\% in
  $v\sin i$; therefore, adoption of a more adequate limb darkening law
  would be a potential improvement.

First, we used synthetic stellar spectra from the Pollux database
\citep{2010A&A...516A..13P} to generate the training set. The model
grid covered $T_\mathrm{eff}$ from \unit[6000]{K} to \unit[15,000]{K}
in steps of \unit[100]{K}, $\log g$ from \unit[3.5]{dex} to
\unit[4.5]{dex} in steps of \unit[0.5]{dex}, and
$\lbrack\mathrm{Fe}/\mathrm{H}\rbrack$ from $\unit[-1.0]{dex}$ to
$\unit[1.0]{dex}$ in steps of \unit[0.5]{dex}. Spectra were computed
based on the plane-parallel ATLAS12 model atmospheres in
LTE \citep{2005MSAIS...8..189K}. The
microturbulent velocity was fixed at $\unit[2]{km\,s^{-1}}$. We used
this original set of ATLAS12 model spectra to train the first SLAM
model and then applied this SLAM model to generate 10,000 synthetic
spectra covering a uniform distribution in $\log(T_\mathrm{eff})$,
$\log g$, and $\lbrack\mathrm{Fe}/\mathrm{H}\rbrack$.

In the second step, the synthetic spectra were convolved with the
rotational broadening and downgraded to a resolution of $R\sim 7500$
to resemble the MRS spectra from both the blue and red arms. We
normalized these mock spectra using the same procedure as used for the
MRS spectra and re-sampled them to a wavelength step of
\unit[0.2]{\AA}, approximately the sampling step of
$\lambda/3R$. Gaussian noise resulting in $\mathrm{SNR} = 40$ was
added to the spectra to simulate the practical conditions pertaining
to the observed spectra.

We adopted a grid of SLAM hyperparameters (parameters used to control
the learning process) defined by $\epsilon = 0.05$, $C = [0.1, 1,
  10]$, and $\gamma = [0.1, 1, 10]$, where $\epsilon$ and $C$
represent the tube radius and penalty level in the SVR algorithm,
respectively, and $\gamma$ represents the width of the kernel of the
radial basis function. These hyperparameters define the complexity of
the SVR model, which is automatically determined by the training set
through the mean-squared error. The best-fitting hyperparameters were
found through five-fold cross-validation, i.e. by partitioning the
entire training sample into five equal-sized subsets and adopting a
single subsample as validation data to test the model trained by the
other four subsamples. This cross-validation was repeated five times,
iterating through all five subsamples.

\begin{figure*}[ht!]
\plotone{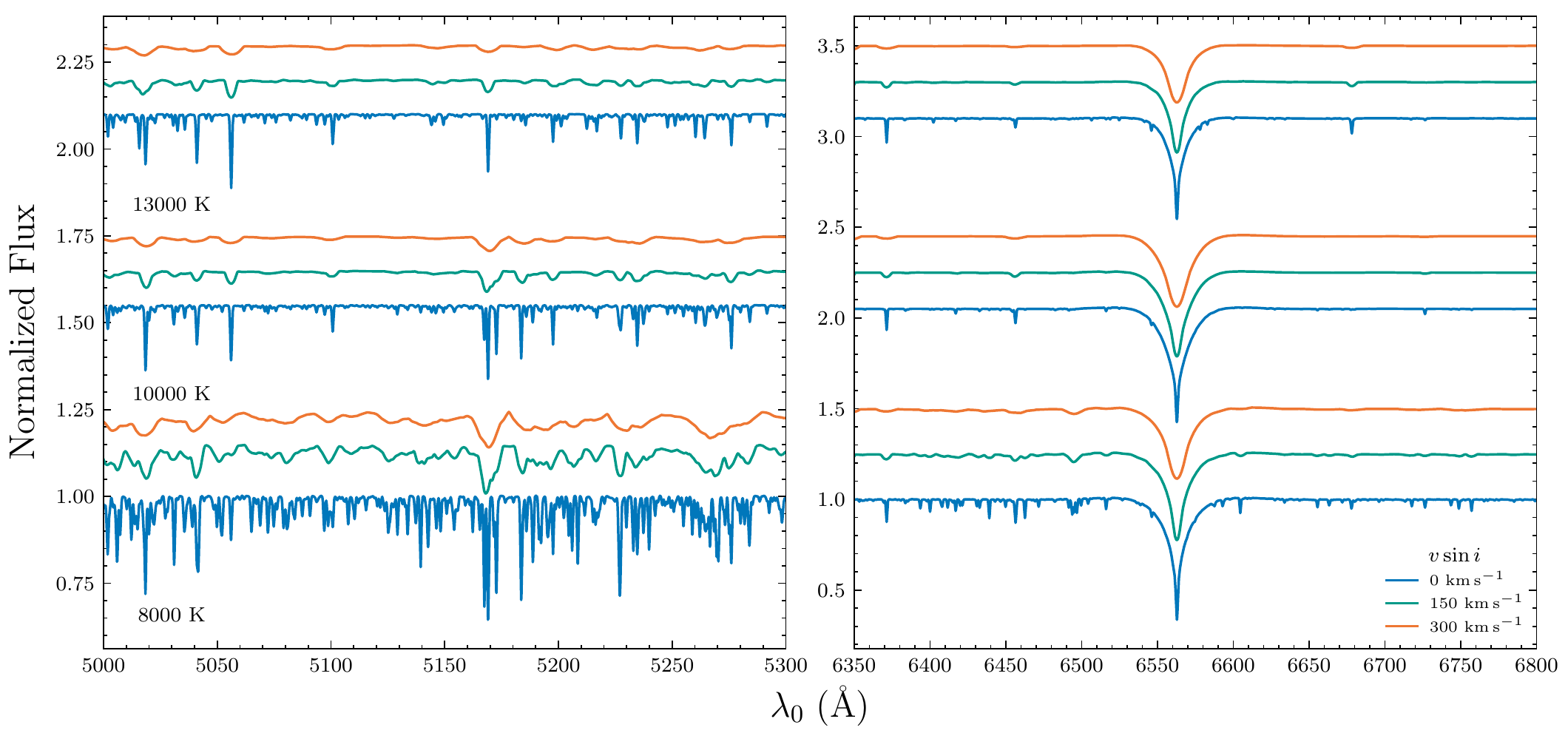}
\caption{Model spectra for $T_\mathrm{eff}=\unit[8000]{K}$
    (bottom), \unit[10000]{K} (middle), and \unit[13000]{K}
    (top). Each group is further offset by $v\sin i =
    \unit[0]{km\,s^{-1}}$ (blue), $\unit[150]{km\,s^{-1}}$ (green),
    and $\unit[300]{km\,s^{-1}}$ (orange). Observations of the same
    object obtained with the blue and red arms are shown in the left
    and right panels, respectively. The model spectra have been
    broadened to the MRS resolving power, $R=7500$, and resampled to
    the MRS wavelength coverage.
\label{fig:model1}}
\end{figure*}

\begin{figure*}[ht!]
\plotone{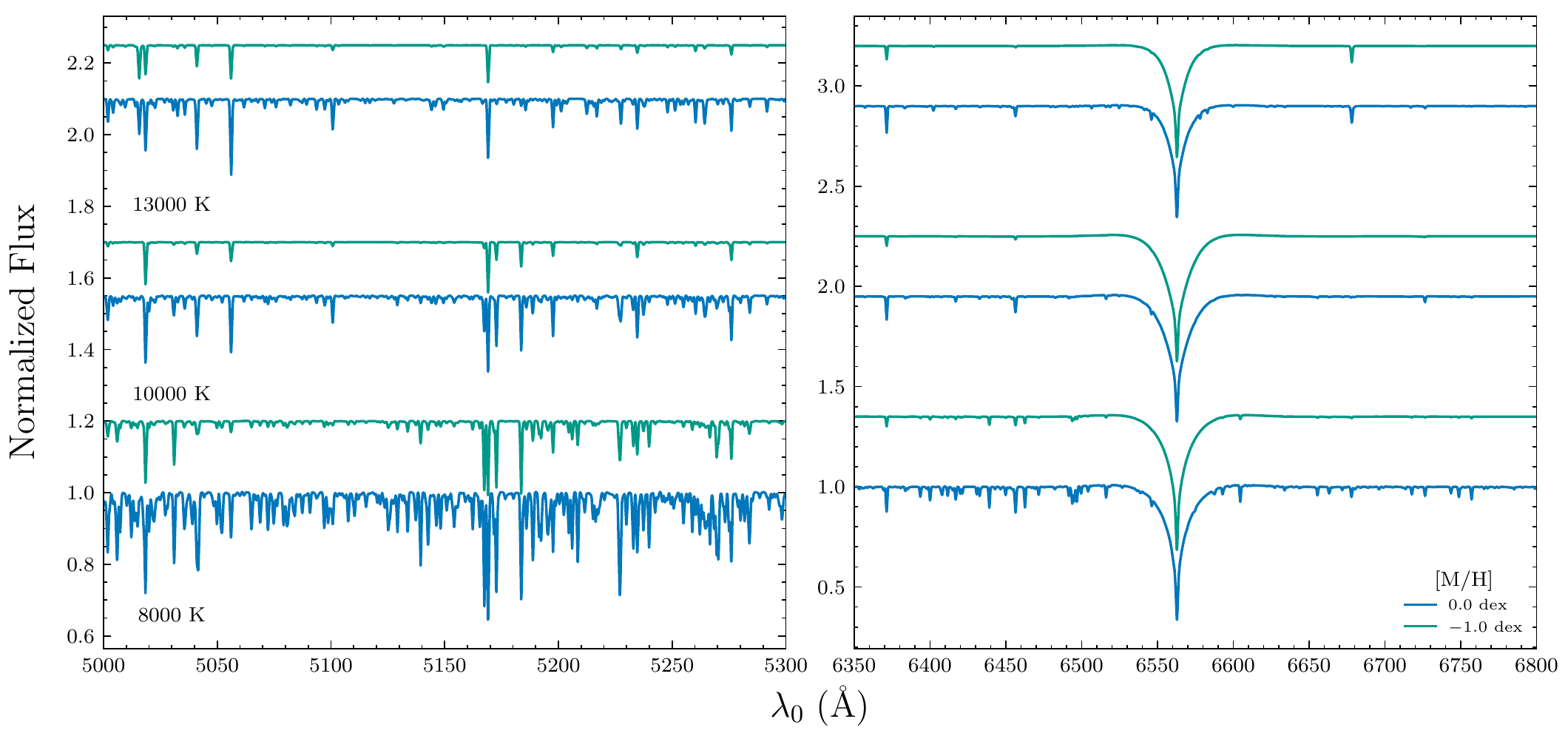}
\caption{As Figure~\ref{fig:model1}, but for different
    metallicities, i.e., $\mathrm{[M/H]}=\unit[0.0]{dex}$ (blue) and
    $\unit[-1.0]{dex}$ (green). \label{fig:model2}}
\end{figure*}

\begin{figure*}[ht!]
\plotone{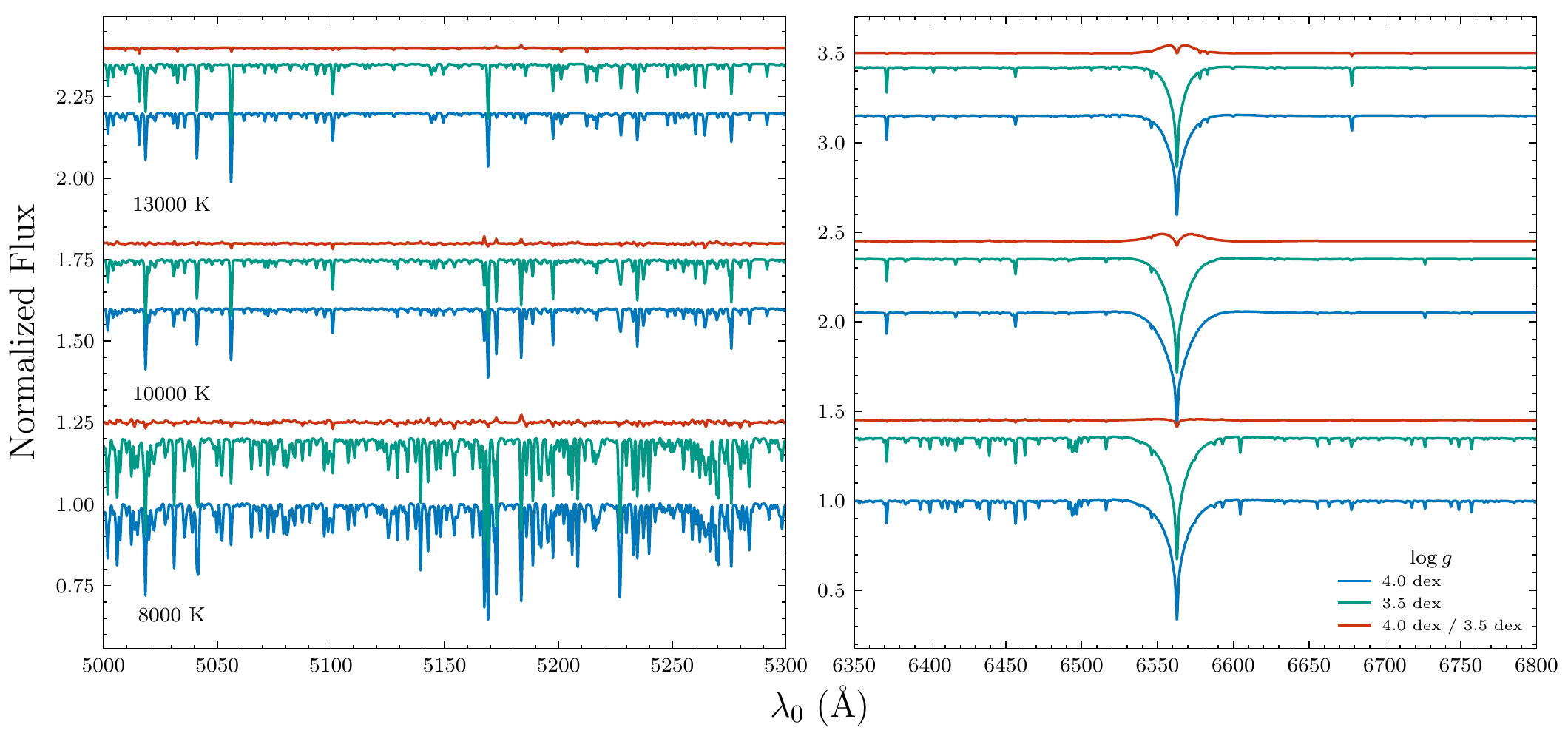}
\caption{As Figure~\ref{fig:model1}, but for different surface
    gravities, i.e., $\log g=\unit[4.0]{dex}$ (blue) and
    $\unit[3.5]{dex}$ (green). The red curve represents the flux ratio
    of the $\log g=\unit[4.0]{dex}$ and $\log g=\unit[3.5]{dex}$
    models, $F_{\log g=4.0}/F_{\log g=3.5}$.\label{fig:model3}}
\end{figure*}

The predicted performance of the SLAM method fundamentally
  depends on how the spectral-line features vary as a function of the
  adopted labels. As one of the first applications to early-type
  stars, we present Figures~\ref{fig:model1}, \ref{fig:model2}, and
  \ref{fig:model3} to illustrate these variations as functions of
  $v\sin i$, [M/H], and $\log g$, respectively. Three groups of model
  spectra for different temperatures are shown at the top
  (\unit[13,000]{K}), middle (\unit[10,000]{K}), and bottom
  (\unit[8000]{K}). The left (right) panels show the blue (red)
  segments of the MRS spectral range. The model spectra were broadened
  to the MRS resolving power, $R=7500$, and resampled to the MRS
  wavelength coverage. The default parameter set includes solar
  metallicity $\mathrm{[M/H]}=0$, $\log g = \unit[4.0]{dex}$, and
  $v\sin i = \unit[0]{km\,s^{-1}}$.

Figure~\ref{fig:model1} and \ref{fig:model2} show clear evidence
  that the model spectra are strongly dependent on the values of
  $v\sin i$ and [M/H]. As the rotation velocity increases, the
  absorption lines become blended and only strong lines remain
  visible. The effects of rotational broadening on the spectra can be
  differentiated easily from the impact of other stellar labels
  ($T_\mathrm{eff}$, [M/H], and $\log g$). This enables us to
  unequivocally probe the effects of stellar rotation. All metal line
  depths decrease with decreasing metallicity, although the H$\alpha$
  profile is largely unchanged (as is He\,{\sc i} 6678 for hotter
  B-type stars). Thus, the potential degeneracy between higher
  temperature and lower metallicity could be alleviated by the
  strength of the H$\alpha$ profile. Moreover, the equivalent widths
  of the Balmer lines are maximal around \unit[9000]{K} due to the
  relatively high excitation energy of the Balmer series
  (\unit[10.2]{eV}) \citep{2005oasp.book.....G}. The presence of
  He\,{\sc i} and/or Si\,{\sc ii}, which only appear at higher
  temperatures, could help to resolve this degeneracy.

Any variations caused by variations in $\log g$ are small, as
  shown in Figure~\ref{fig:model3}. To provide to a clearer picture of
  the latter effect, the flux ratios of the $\log g=\unit[4.0]{dex}$
  and $\log g=\unit[3.5]{dex}$ models, $F_{\log g=4.0}/F_{\log
    g=3.5}$, are overplotted as red curves. Generally, the dependence
  on $\log g$ is weak. In the blue arm, mostly only the Mg\,{\sc i}
  \textit{b} lines exhibit some changes for cooler models. As for
  H$\alpha$ in the red arms, its wings are merely influenced by Stark
  broadening for the hotter models with lower gravity, and only the
  line cores show some dependence on $\log g$. This suggests a poor
  capability for predicting the surface gravity, particularly for
  cooler stars.

\subsection{Uncertainties}

We will discuss two types of uncertainties, internal errors associated
with the SLAM method itself and external uncertainties associated with
the observed spectra.

The internal uncertainty pertaining to the optimization method could
be represented by the SLAM errors, as converted from the Hessian
matrix. The diagonal elements of the covariance matrix were used as
the formal errors. We also estimated the precision of the model based
on the cross-validated scatter ($\sigma_\mathrm{CV}$), which is the
standard deviation of the difference between the predicted and true
values:
\begin{equation}
	\sigma_\mathrm{CV} =
\frac
{1}{m}\sqrt{\sum_{i=1}^m\left(\theta_{i,\mathrm{SLAM}}-\theta_i\right)^2},
\end{equation}
where $\theta_i$ and $\theta_{i,\mathrm{SLAM}}$ are the true and
predicted stellar labels, respectively. The cross-validated
  scatter was calculated based on a subset composed of one-tenth of
  the full training sample. This subset was randomly selected to
  ensure uniformity across the parameter ranges.

\begin{figure*}[ht!]
\plotone{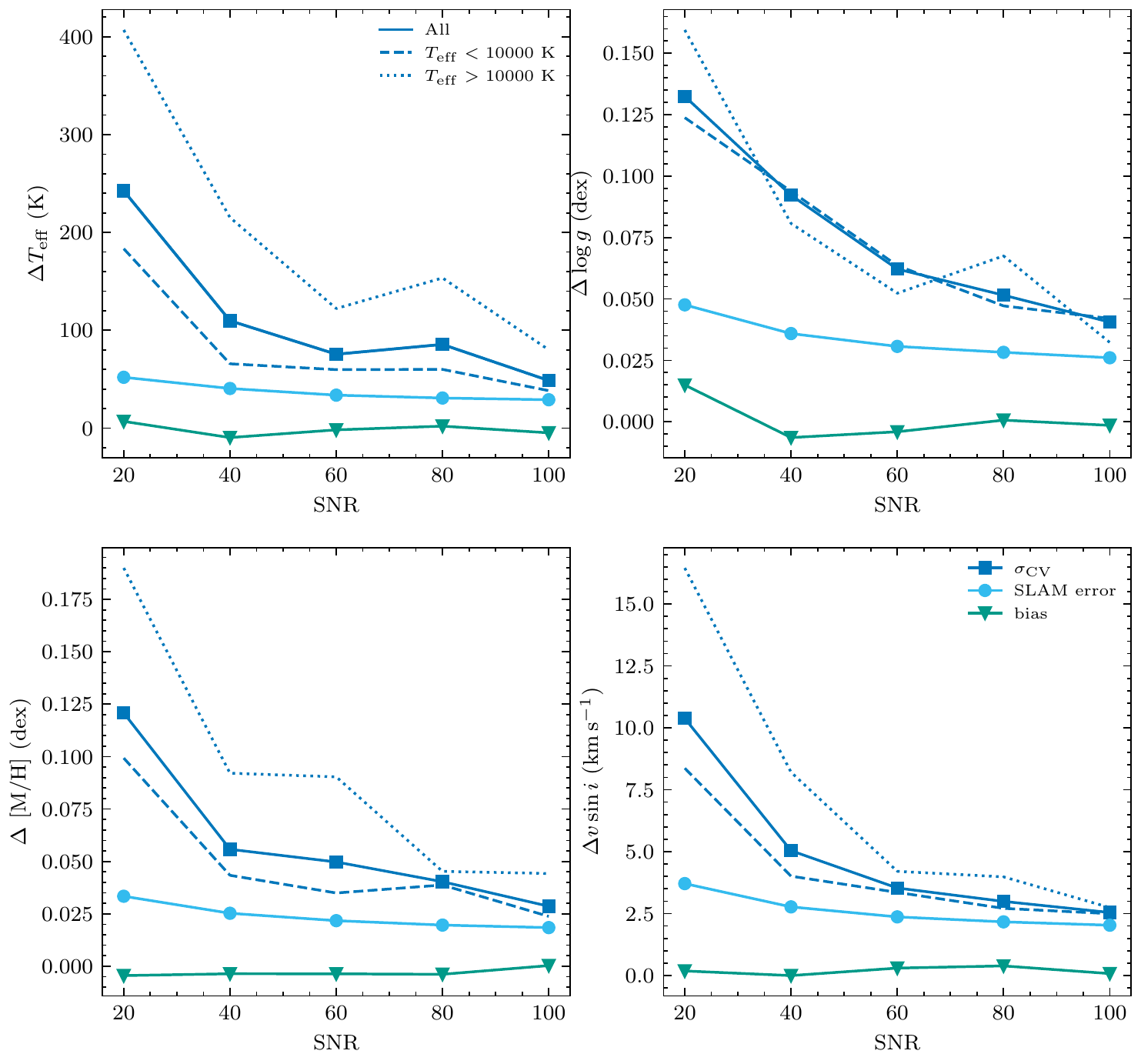}
\caption{Uncertainties in the stellar labels ($T_\mathrm{eff}$, $\log
  g$, [M/H], and $v\sin i$) versus SNR. Blue squares, cyan
  circles, and green triangles represent the cross-validated scatter,
  $\sigma_\mathrm{CV}$, SLAM error, and bias, respectively. The
  $\sigma_\mathrm{CV}$ values of $T_\mathrm{eff}$, $\log g$,
    [M/H], and $v\sin i$ mostly converge at $\sim\unit[75]{K}$,
  \unit[0.06]{dex}, \unit[0.05]{dex}, and
  $\sim\unit[3.5]{km\,s^{-1}}$, respectively, for $\mathrm{SNR} >
  60$. In each panel, the $\sigma_\mathrm{CV}$ values of
  the corresponding label for two different subsamples, split by
  their temperatures, are shown as dashed ($T_\mathrm{eff} <
  \unit[10,000]{K}$) and dotted ($T_\mathrm{eff} > \unit[10,000]{K}$)
  lines.\label{fig:snr}}
\end{figure*}

Figure~\ref{fig:snr} shows the uncertainties in the stellar labels for
different SNRs, from 20 to 100. For objects with $\mathrm{SNR} > 60$,
the $\sigma_\mathrm{CV}$ values for $T_\mathrm{eff}$, $\log g$,
  [M/H], and $v\sin i$ are $\sim\unit[75]{K}$, \unit[0.06]{dex},
\unit[0.05]{dex}, and $\sim\unit[3.5]{km\,s^{-1}}$, respectively. Both
$\sigma_\mathrm{CV}$ and the SLAM error converge for $\mathrm{SNR} >
60$, but $\sigma_\mathrm{CV}$ is much larger than the SLAM error. This
is due to the unconsidered errors in the stellar labels in both the
training and validation processes. Thus, following
\citet{2020ApJS..246....9Z}, we recommend the cross-validated scatter
as a more reliable estimate to evaluate the performance of the SLAM
model. The cross-validated test also yields the biases (see the
triangles in Figure~\ref{fig:snr}), which contribute less than 10\% to
the overall uncertainties.

Note that the uncertainties just discussed represent the average
behavior of the SLAM model. Its performance will vary across a wide
range of the relevant parameter space, especially for
$T_\mathrm{eff}$. The top left panel of Figure~\ref{fig:snr} shows
$\sigma_\mathrm{CV}$ for two different subsamples, divided by their
temperatures, $T_\mathrm{eff} < \unit[10,000]{K}$ (dashed line) and
$T_\mathrm{eff} > \unit[10,000]{K}$ (dotted line). The performance of
the hotter subsample is much worse than that of the cooler stars.
  The same trend with effective temperature is seen for [M/H] and
  $v\sin i$. Figure~\ref{fig:model1} shows that the available metallic
  absorption features become weaker as the temperature increases,
  particularly for the blue arm. This lack of information leads to
  larger uncertainties in estimates of the stellar labels for hotter
  stars, which also holds for lower metallicities. The exception is
  $\log g$ (top right), where there is no significant difference
  between the $\sigma_\mathrm{CV}$ values for the hotter and cooler
  subsamples. This could be due to our use of both the blue and red
  segments of the MRS spectra, which broadly offsets the spectral
  dependence on effective temperature.

\subsection{Validation}

\begin{figure*}[ht!]
\plotone{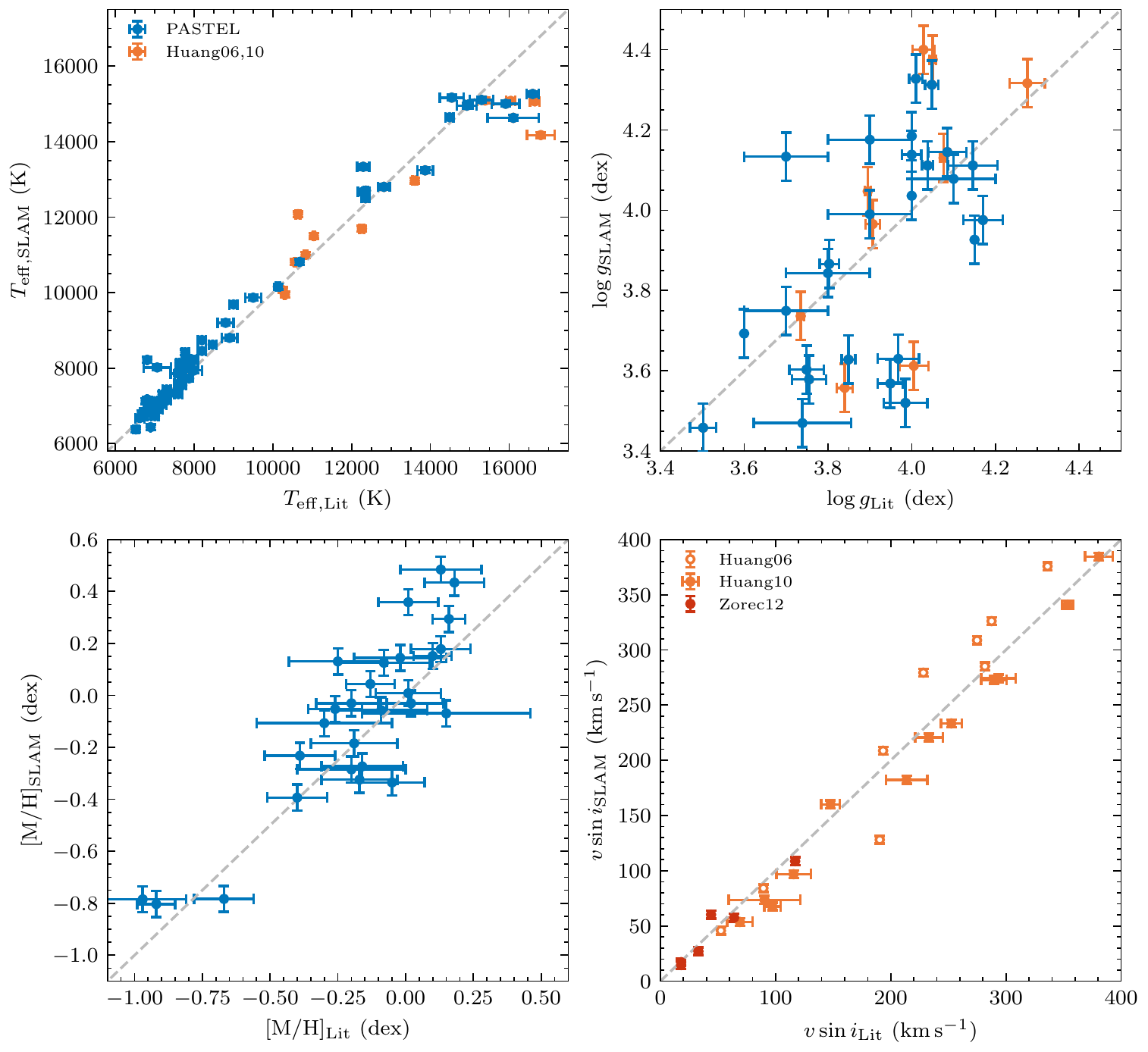}
\caption{Comparison of stellar labels from SLAM and literature results
  \citep{2006ApJ...648..591H, 2006ApJ...648..580H,
    2010ApJ...722..605H, 2012A&A...537A.120Z,
    2016A&A...591A.118S}. (Top left) $T_\mathrm{eff}$ shows great
  consistency for $T_\mathrm{eff} < \unit[15,000]{K}$, the upper limit
  from the training model. The vertical error bars are on the order of
  or smaller than the data points. (Top right) $\log g$ from SLAM
  marginally agrees with the literature values, with significant
  scatter. (Bottom left) [M/H] from SLAM follows the one-to-one
  correlation with the literature results. (Bottom right) Our
  estimates of $v\sin i$ are consistent with the one-to-one relation
  with the literature values. The $v\sin i$ values from
  \citet{2006ApJ...648..591H} and \citet{2010ApJ...722..605H} were
  calibrated based on high-precision measurements. The error bars for
  our estimates (horizontal axis) come from the cross-validated
  scatter ($\sigma_\mathrm{CV}$).
\label{fig:verify}}
\end{figure*}

A comparison of the stellar labels derived from SLAM and those
collected from the literature is shown in Figure~\ref{fig:verify}. The
validation sample comes from \citet{2006ApJ...648..591H,
  2006ApJ...648..580H}, \citet{2010ApJ...722..605H},
\citet{2012A&A...537A.120Z}, and
\citet{2016A&A...591A.118S}. \citet{2016A&A...591A.118S} is a
bibliographical catalog containing determinations of stellar
atmospheric parameters. Its stellar parameters ($T_\mathrm{eff}$,
$\log g$, and [M/H]) were collected from the literature based on
high-resolution ($R\geqslant 25,000$) and high SNR (SNR $> 50$)
spectra. Although these parameters were not derived homogeneously, the
median value of the standard deviation of the values from different
measures is 0.76\%. The main sources of rotational velocities are
\citet{2006ApJ...648..580H} and \citet{2010ApJ...722..605H}. Those
authors conducted a survey of B-type stars based on
moderate-resolution ($R\sim 2000$--4000) spectra from the WIYN
\unit[3.5]{m} and CTIO \unit[4]{m} telescopes.

The top left panel of Figure~\ref{fig:verify} presents a comparison of
the effective temperatures. Because \citet{2012A&A...537A.120Z}
estimated their temperatures based on Str{\"o}mgren photometric color
indices, which is inconsistent with the spectroscopic method used for
our sample and other literature sources, we did not include their
estimates. There is great consistency between our results and the
literature results up to $T_\mathrm{eff} = \unit[15,000]{K}$. This is
the temperature limit inherited from the training model, where the
upper limit is set by the parameter grid of the atmospheric
models. Thus, we disregard $T_\mathrm{eff}$ comparisons beyond this
limit.

As for $\log g$ (see the top right panel of Figure~\ref{fig:verify}),
it demonstrates the worst performance among the four stellar labels:
$\log g$ from SLAM roughly agrees with the literature values, although
with significant scatter. This poor performance of $\log g$ has been
investigated by \citet{2020RAA....20...51Z}, who used the Coefficient
of Dependence (COD) to quantify the global spectral information
content. They found that the COD of $\log g$ remains low for dwarf
stars. This means that the spectral information content in $\log g$ is
not well carried in the MRS spectra. This is, however, not a major
issue for our analysis as the resulting $v\sin i$ values are not
significantly affected by this poor performance (as demonstrated in
the bottom right panel of Figure~\ref{fig:verify}). Our analysis of
the bolometric luminosities (see Section~\ref{sec:bol}) and, thus, the
stellar masses (see Section~\ref{sec:mass}) depends only weakly on
this stellar label.

[M/H] from SLAM follows the one-to-one correlation with the
literature results. However, there might be an offset of
\unit[0.08]{dex} toward higher metallicity. This might be due to the
intrinsic error in the training model. However, the linear correlation
with a slope close to unity ensures the reliability of our estimates.

The bottom right panel of Figure~\ref{fig:verify} presents a
comparison of $v\sin i$ values. Despite their low resolution ($R\sim
2000$--4000), \citet{2006ApJ...648..591H} and
\citet{2010ApJ...722..605H} achieved sufficient SNR ($> 50$) to ensure
the reliability of their measurements. Their results have been
validated by \citet{2012A&A...537A.120Z}, yielding an almost
one-to-one correlation (with a slope of $0.971\pm0.04$) with the
results from high-resolution ($R> 20,000$) spectra. For the $v\sin i$
comparison, we used the calibration function derived by \citet[][their
  Equation 3]{2012A&A...537A.120Z} to correct their
values. \citet{2012A&A...537A.120Z} compiled the current-largest
catalog of 2014 A-type field stars, which includes
\citet{2002A&A...381..105R, 2002A&A...393..897R}'s results. However,
because our sample is on average $\sim\unit[4]{mag}$ fainter than
theirs, there are only six matches (included in
Figure~\ref{fig:verify}). Linear regression yields,
\begin{equation}
	v\sin i_\mathrm{SLAM} = 1.05_{\pm 0.04} v\sin i_\mathrm{Lit} - 4.97_{\pm
8.37},
\label{eq:calibrate}
\end{equation}
which is a remarkable consistency in measures of $v\sin i$. The offset
is not significant compared with its uncertainty, which could be
associated with the variable instrumental broadening. The actual
spectral resolution of a fiber depends on its location on the
observation plate, which may generate slightly different broadening
values. The wavelength coverage of a given spectrum is also affected
by the fiber positioning, leading to variations in the information
retained in the data. Both effects introduce possible biases in
estimates of stellar rotation, but they are well-constrained in the
slow rotation regime.

Ideally, rotation measurements should be linked to surface gravity
measurements, in the sense that the surface gravity determines the
steepness of the pressure gradient, which in turn generates the
pressure broadening, a second broadening mechanism in addition to the
rotational instrumental broadening. The uncertainties in estimating
$\log g$ should therefore affect the $v\sin i$ derivation. In our
case, this effect could be mediated in one of three ways. First,
pressure broadening strongly affects predominantly the Balmer lines
rather than the narrow lines (e.g. Mg\,{\sc i} \textit{b});
rotational information is well preserved if one were to use
narrow-line absorption. Second, the median-resolution spectrograph
enables us to resolve the different broadened rotation and pressure
profiles. Using a method similar to profile fitting rather than
calculating the full width at half maximum, we can disentangle these
two mechanisms. Third, the vast majority of our sample objects are
dwarf stars. They span a relatively small range in $\log g$, around
\unit[4.0]{dex}, compared with the giant stars. Thus, the impact of
large uncertainties in $\log g$ can be controlled adequately.

Moreover, the consistency between our estimates and the literature
results can extend to lower $v\sin i$, even below the detection limit
for MRS resolution, suggesting that our estimates can potentially
apply also to very slow rotators. Since the offset (if any) is minor,
we used the results from the SLAM model without adjustments. For
further use of complementary samples, we recommend using the inverse
function of Equation~\ref{eq:calibrate} for calibration.

\section{Results}
\label{sec:results}

Here we discuss our procedures to derive the fundamental parameters
(bolometric luminosity, mass, and age) from the SLAM stellar labels,
including the extinction and bolometric corrections
(Section~\ref{sec:bol}), and the use of the stellar evolutionary
diagram (Section~\ref{sec:mass}). We further demonstrate the impact of
our selection criteria on these fundamental parameters, as well as the
stellar labels, allowing us to derive the final catalog.

\subsection{Bolometric luminosities}
\label{sec:bol}

Since the majority of our sample is embedded in the Galactic disk, we
use the high-precision photometry from \textit{Gaia}, combined with
2MASS photometry given that it is only marginally affected by
extinction, to estimate the bolometric luminosities. We infer the
extinction toward each star based on the difference between the
observed \textit{Gaia} and the intrinsic color indices. The intrinsic
color index, $\left(G_\mathrm{BP} - G_\mathrm{RP}\right)_0$, was
derived from the PARSEC 1.2S \citep{2012MNRAS.427..127B} evolutionary
tracks by interpolating the model grid defined by $\log
T_\mathrm{eff}$, $\log g$, and $\lbrack\mathrm{M}/\mathrm{H}\rbrack$
(assuming no $\alpha$ enhancement) to the corresponding values derived
from SLAM.

\begin{figure}[ht!]
\plotone{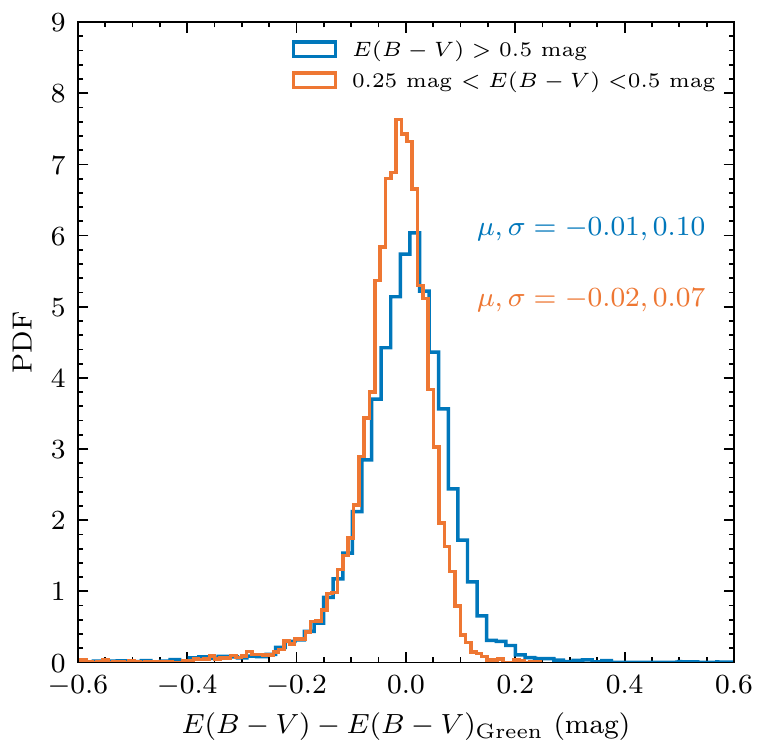}
\caption{Distribution of the difference in $E(B-V)$ estimated in this
  paper and by \citet{2019ApJ...887...93G}. The comparisons are for
  stars with $E(B-V)>\unit[0.5]{mag}$ (blue) and
  $\unit[0.25]{mag}<E(B-V)<\unit[0.5]{mag}$ (orange). The mean values
  of the differences ($\mu$) and the dispersions ($\sigma$) are also
  shown.\label{fig:extinction}}
\end{figure}

The typical uncertainty in $E(G_\mathrm{BP}-G_\mathrm{RP})$ should be
on the order of \unit[0.1]{mag}
\citep[e.g.][]{2019ApJ...887...93G}. Figure~\ref{fig:extinction} shows
a comparison of $E(B-V)$ values estimated in this paper and by
\citet{2019ApJ...887...93G}. The transformation from measured
  $E(G_\mathrm{BP}-G_\mathrm{RP})$ to $E(B-V)$ was calculated based on
  the \citet{1989ApJ...345..245C} and \citet{ 1994ApJ...422..158O}
  extinction law with $R_\mathrm{V}=3.1$. The extinction from
  \citet{2019ApJ...887...93G} was queried based on dust maps
  \citep{2018JOSS....3..695G} and converted to $E(B-V)_\mathrm{Green}$
  using Table 6 of \citet{2011ApJ...737..103S}.

Since our sample stars are mostly
embedded in the Galactic disk, we selected two subsamples affected by
significant reddening, i.e. $E(B-V)>\unit[0.5]{mag}$ and
$\unit[0.25]{mag}<E(B-V)<\unit[0.5]{mag}$. Comparison with the
\citet{2019ApJ...887...93G} reddening map for high extinction suggests
a typical uncertainty of $\sim\unit[0.1]{mag}$ in $E(B-V)$. There is
great consistency between both estimates, although our results might
be biased toward lower extinction by $\sim\unit[0.01]{mag}$.

The bolometric correction (BC) was adopted from
\citet{2019A&A...632A.105C}, who considered the effects of spectral
type and non-linearities. The bolometric luminosities were calculated
as
\begin{equation}
	M_\mathrm{bol} = m_{K_\mathrm{S}} + BC_{K_\mathrm{S}} - \mathrm{DM} -
\frac{A_{K_\mathrm{S}}}{E({G_\mathrm{BP}}-{G_\mathrm{RP}})}E(G_\mathrm{BP}-G
_\mathrm{RP}),
\label{eq:bol}
\end{equation}
where $\mathrm{DM}$ is the distance modulus calculated from the
\textit{Gaia} EDR3 parallaxes \citep{2021AJ....161..147B} and $A_i$ is
the extinction coefficient for a given passband, $i$, based on
\citet{1994ApJ...422..158O}. We adopted the 2MASS $K_\mathrm{s}$ band
to mitigate the influence of extinction in the Galactic plane. Since
the photometric precision of the 2MASS $K_\mathrm{s}$ band is better
than \unit[0.03]{mag} for our sample \citep{2006AJ....131.1163S} and
infrared bands are less affected by extinction
($\frac{A_{K_\mathrm{s}}}{E({G_\mathrm{BP}}-{G_\mathrm{RP}})} =
0.26$), the choice of $K_\mathrm{s}$ leads to better estimates of
$M_\mathrm{bol}$. The bolometric luminosities were derived by adopting
$M_\mathrm{\odot, bol}=\unit[4.74]{mag}$. The corresponding error was
estimated by bootstrap sampling of all determinations using
Equation~\ref{eq:bol}, including the stellar labels. The median value
of the uncertainties in the bolometric luminosities is $\Delta \log
(L/L_\odot)=0.02$.

\begin{figure}[ht!]
\plotone{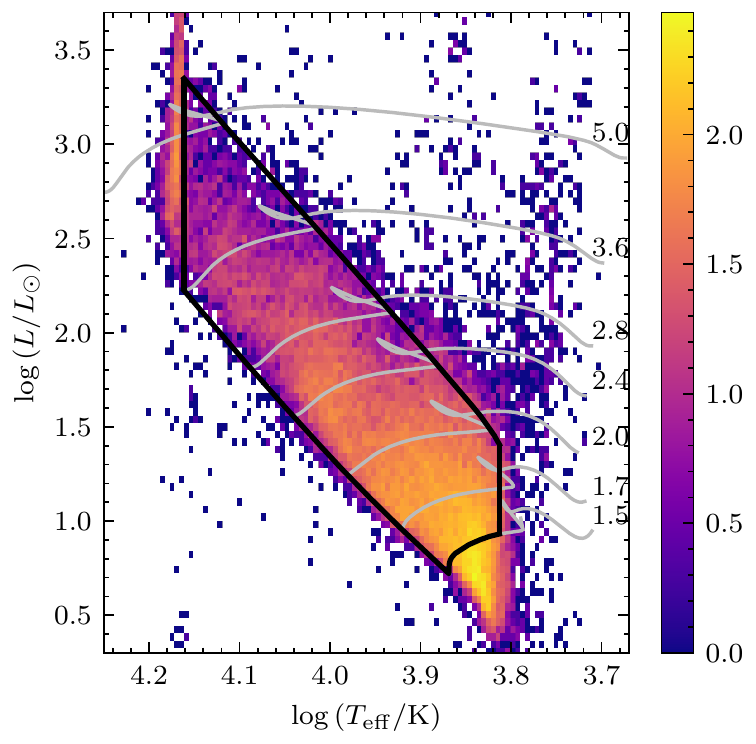}
\caption{Hertzsprung--Russell diagram of single, normal
  stars. Evolutionary tracks of $\unit[1.5, 1.7, 2.0, 2.4, 2.8, 3.6,
    5]{M_\odot}$ are shown as gray curves. Black reference lines outline
  the selection boundary for stars of solar metallicity. The colors
  represent the logarithm of the stellar numbers in each
  bin.\label{fig:lt}}
\end{figure}

We show the Hertzsprung--Russell diagram of our sample objects in
Figure~\ref{fig:lt}. The majority of the sample have effective
temperatures between \unit[6500]{K} and \unit[15,000]{K}. The lower
boundary reflects the rejection of stars with $T_\mathrm{eff} <
\unit[6500]{K}$---$\log(T_\mathrm{eff}/\unit{K}) < 3.81$---in
Section~\ref{sec:cat}, whereas the upper boundary is limited by the
hottest stars in the ATLAS atmospheric models. Note that there is an
abnormal concentration close to the \unit[15,000]{K} boundary. This is
an artifact caused by the temperature limit in the training
process. Thus, we reject all stars with $T_\mathrm{eff} >
\unit[14,500]{K}$---$\log(T_\mathrm{eff}/\unit{K}) > 4.16$---(black
vertical line) from further analysis. The aggregation around
\unit[9400]{K}---$\log(T_\mathrm{eff}/\unit{K}) \sim 3.97$---shown in
Figure~\ref{fig:ew} no longer exists, suggesting that it does not
reflect any real physical property.

In addition to a temperature cut, we also limited our selection to
those stars that are still on the main sequence. As stars evolve off
the core-hydrogen-burning phase, their rotation velocities decelerate
significantly owing to the conservation of angular momentum during the
evolutionary expansion of subgiant-branch stars
\citep[e.g.,][]{2016ApJ...826L..14W}. Because these evolved stars may have
$\log g$ beyond
our model grids, we cannot confirm the accuracy of any resulting
estimates. Our selection boundary also includes some stars in the
small contraction phase before they move toward the red giant
branch. This phase is a rapid transition from core to shell hydrogen
burning. Therefore, any contamination associated with this phase will
be of minor importance.

\subsection{Stellar masses and ages}
\label{sec:mass}

\begin{figure}[ht!]
\plotone{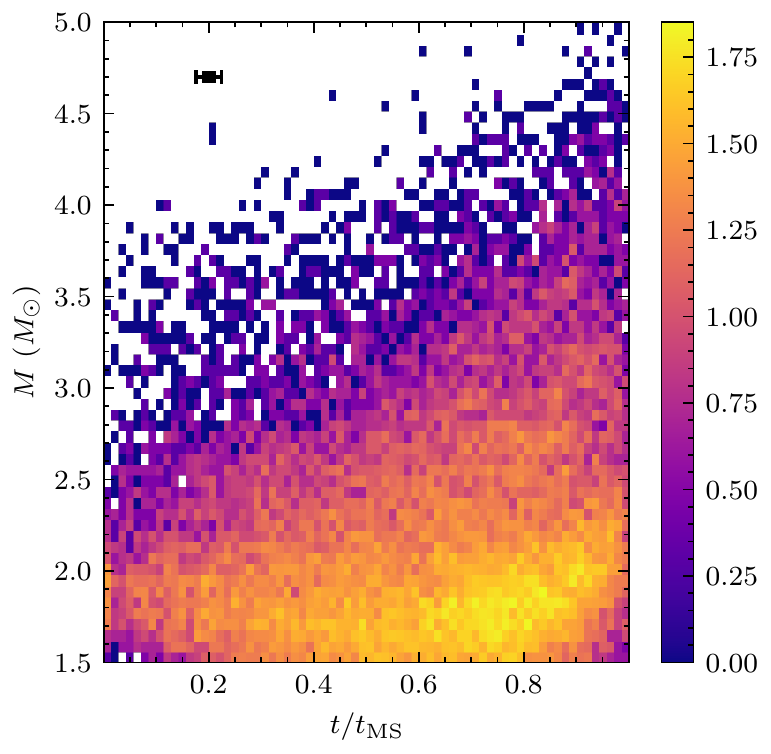}
\caption{Mass and age ($t/t_\mathrm{MS}$) distributions of selected
  single, normal stars. Typical errors are indicated by the error bar
  at top left. \label{fig:mass}}
\end{figure}

We used PARSEC 1.2S evolutionary tracks in the $\log
T_\mathrm{eff}$--$\log L$ space to estimate individual stellar masses
and ages for a given metallicity, [M/H]. Following
\citet{2012A&A...537A.120Z}, we represent the stellar age by $t$, for
the age in years, as well as by $t/t_\mathrm{MS}$, for the life span
from the ZAMS to the terminal-age main sequence (TAMS). The
corresponding error was estimated by resampling of $\log
T_\mathrm{eff}$ and $\log L$. Figure~\ref{fig:lt} shows that
  parts of the evolutionary tracks of the shell H-burning phase reside
  within our selection boundary for main-sequence stars, generating a
  degeneracy between mass and age. In order to test its influence on
  our approach, we performed the same analysis on a mock data set,
  extracted from theoretical evolutionary tracks. We verified that the
  degeneracy is only important for objects at the very end of the
  evolutionary phase ($t/t_\mathrm{MS}>0.95$). The corresponding error
  is smaller than 2\% for both the resulting mass and age estimates.

In Figure~\ref{fig:mass}, we present the distribution in the
$t/t_\mathrm{MS}$--$M$ plane of the single, normal stars selected
within the boundary defined in Figure~\ref{fig:lt}. The general
distribution of our sample in the mass--age diagram displays a similar
profile as that of \citet{2012A&A...537A.120Z}. The trend toward
  larger $t/t_\mathrm{MS}$ with increasing mass reflects the
  temperature cut at $T_\mathrm{eff}=\unit[14500]{K}$ (the vertical
  line in Figure~\ref{fig:lt}). However, our sample fills the lack
of objects observed for the lower mass regime (their Figure~4).

Fast rotators ($v\sin i > \unit[300]{km\,s^{-1}}$) will change their
surface temperatures and luminosities significantly compared with
their non-rotating counterparts. This is mainly due to (1) the notion
that the centrifugal support results in a decrease in the surface
temperature and luminosity by decreasing the effective gravity
\citep{1997A&A...321..465M}; and (2) the orientation of a star with
respect to the line of sight determines the observed fraction of the
stellar surface \citep{1985MNRAS.213..519C}. Such mechanisms are not
included in the \citet{2012MNRAS.427..127B} non-rotating evolutionary
tracks and might introduce additional biases and uncertainties into
estimates of fundamental stellar parameters. This caveat could be
alleviated by employing detailed asteroseismic modeling
\citep[e.g.][]{2017NatAs...1E..64C,
  2019MNRAS.487..782L}. Unfortunately, such information is not
available for our sample. Nevertheless, parameters based on
non-rotating models are slightly influenced by rotation
effects. \citet{2012A&A...537A.120Z} compared the masses and ages
determined based on non-rotating evolutionary tracks with those
resulting from rotation \citep{2000A&A...361..101M,
  2008A&A...478..467E, 2011A&A...530A.115B, 2012A&A...537A.146E}. They found
that
differences are smaller than or around 10\% if the rotation is not
close to the critical rotation rate ($\Omega/\Omega_\mathrm{crit}$,
where $\Omega$ is the angular rotation velocity). Since only a
fraction of our sample objects are rapid rotators, our implementation
of non-rotating evolutionary tracks is reasonable.

As our final selection step, we discarded any star with large errors
in either mass or relative age ($\sigma_M \geq \unit[0.1]{M_\odot}$,
$\sigma_{t/t_\mathrm{MS}} \geq 0.15$). To ensure the accuracy of our
$v\sin i$ estimates, we also made a selection based on the associated
SLAM error ($\sigma_{v\sin i}/v\sin i < 0.05$). Combined with our
temperature selection (Section~\ref{sec:bol}), the final catalog
contains 40,034 stars. All stellar labels and fundamental parameters
are listed in Table~\ref{tab:descript}.

\begin{deluxetable*}{lllp{100mm}}
\tablecaption{Contents of the Catalog\label{tab:descript}}
\tablewidth{0pt}
\tablehead{\colhead{Num}&\colhead{Column}&\colhead{Unit}&\colhead{Description}}
\startdata 
1     & ID                                 &           & \textit{Gaia} DR3
source ID                    \\
2     & obsid                                 &           & LAMOST MRS
observation ID                    \\
3     & RA                                 &   \degr   & Right Ascension
(J2000)                    \\
4     & Dec                                &   \degr   & Declination (J2000)   
\\
5, 6  & $\mathrm{EW}_\mathrm{H\alpha}$, $\mathrm{EW}_\mathrm{Mg}$ & \AA & Line
indices of H$\alpha$ and Mg\,{\sc i} \textit{b} EW \\
7, 8  & $v_\mathrm{R,b}$, $v_\mathrm{R,r}$                           &
$\unit{km\,s^{-1}}$       & Radial velocities, blue and red arms               
\\
9     & $T_\mathrm{eff}$                    & K         & Effective temperature
from SLAM\\
10    & $\sigma_{T_\mathrm{eff}}$                            &   K        &
Uncertainty in $T_\mathrm{eff}$ (SLAM error)              \\
11    & $\log g$                                &   dex        & Surface
gravity from SLAM\\
12    & $\sigma_{\log g}$                                &   dex        &
Uncertainty in $\log g$ (SLAM error)             \\
13    & [M/H]                          &   dex        & Metallicity from
SLAM 
\\
14    & $\sigma_\mathrm{[M/H]}$                           &    dex      
&
Uncertainty in [M/H]  (SLAM error)     \\
15    & $v\sin i$                                  &    $\unit{km\,s^{-1}}$    
& Projected rotation velocity from SLAM                     \\
16     & $\sigma_{v\sin i}$                                  &  
$\unit{km\,s^{-1}}$        & Uncertainty in $v\sin i$ (SLAM error)             
\\
17     & $\log L/L_\odot$                                  &           &
Logarithmic luminosity             \\
18     & $\sigma_{\log L}$                                  &           &
Uncertainty in $\log L/L_\odot$    \\
19     & $M$ &    $M_\odot$       & Stellar mass             \\
20     & $\sigma_{M}$ &      $M_\odot$     & Uncertainty in $M$           \\
21 & $t/t_\mathrm{MS}$                       &  & Life span from the ZAMS
to the TAMS                 \\
22 & $\sigma_{t/t_\mathrm{MS}}$                       &  & Uncertainty in
$t/t_\mathrm{MS}$                  \\
23, 24 & $\mathrm{SNR_b}$, $\mathrm{SNR_r}$ & & Signal-to-noise ratio for
blue
and red arms \\
25, 26 & $\chi^2_\mathrm{b}$, $\chi^2_\mathrm{r}$ & & $\chi^2$
estimates of the
best fit of the blue and red arms
\enddata
\tablenotetext{}{(Only a portion of this table is shown here to demonstrate its
form and content. This table is available in its entirety in machine-readable
form in the online journal.)}
\end{deluxetable*}

\section{Conclusions}
\label{sec:conclusions}

We have compiled a spectroscopic catalog of stellar parameters for
40,034 early-type stars with $\unit[7000]{K} \leq T_\mathrm{eff} \leq
\unit[14,500]{K}$. The primary selection of the LAMOST DR7 MRS spectra
was carried out by considering the line indices of H$\alpha$ and
Mg\,{\sc i} \textit{b}. We used a synthetic atmospheric model library
to build the training set for a data-driven model to estimate the
stellar labels. The precisions of our estimates are
$\sim\unit[75]{K}$, \unit[0.06]{dex}, \unit[0.05]{dex}, and
$\sim\unit[3.5]{km^{-1}}$ for $T_\mathrm{eff}$, $\log g$, [M/H],
and $v\sin i$, respectively, for $\mathrm{SNR} > 60$.

As the first in a series of papers aimed at exploring stellar
rotation, we have illustrated the large volume of our sample and the
capability of MRS spectra to sample stellar projected rotation
  velocities as small as a few $\unit{km\,s^{-1}}$. This will help
answer fundamental questions regarding the origins and characteristics
of their angular momentum. Note that the rotation behavior of our
sample might be subject to external effects, e.g. binary and
chemically peculiar stars. Cluster members may have a different
rotation distribution compared with the field
\citep{2006ApJ...648..580H, 2019ApJ...883..182S}. These
  extenuating factors might influence analyses of stellar
rotation. We will address this issue in Paper II, where we discuss our
sample's statistical properties. Further studies should quantitatively
investigate rotation as a function of stellar mass, age, metallicity,
etc. Finally, a better theoretical characterization of the angular
momentum of early-type stars in our sample would be beneficial to
better understand how to recover the rotation distribution.

\acknowledgments L. D. acknowledges research support from the National
Natural Science Foundation of China through grants 11633005, 11473037,
and U1631102. The Guoshoujing Telescope (the Large Sky Area
Multi-Object Fiber Spectroscopic Telescope; LAMOST) is a National
Major Scientific Project built by the Chinese Academy of
Sciences. Funding for the project has been provided by the National
Development and Reform Commission. LAMOST is operated and managed by
the National Astronomical Observatories, Chinese Academy of
Sciences. This work has made use of data from the European Space
Agency (ESA) mission \textit{Gaia}
(\url{https://www.cosmos.esa.int/gaia}), processed by the
\textit{Gaia} Data Processing and Analysis Consortium (DPAC;
\url{https://www.cosmos.esa.int/web/gaia/dpac/consortium}). Funding
for the DPAC has been provided by national institutions, in particular
the institutions participating in the \textit{Gaia} Multilateral
Agreement.

\vspace{5mm}
\facilities{LAMOST}
\software{PARSEC \citep[1.2S;][]{2012MNRAS.427..127B}, Astropy
  \citep{2013A&A...558A..33A}, Matplotlib \citep{2007CSE.....9...90H},
  TOPCAT \citep{2005ASPC..347...29T}, dustmaps \citep{2018JOSS....3..695G}}

\end{document}